\documentclass[10pt,twocolumn,letterpaper]{article}

\usepackage{iccv}              %

\newcommand{\vix}{\checkmark\kern-1.2ex\raisebox{0.7ex}{\rotatebox[origin=c]{125}{\textbf{--}}}}

\usepackage{multirow}
\usepackage[abs]{overpic}
\usepackage{adjustbox}
\usepackage{makecell}
\usepackage{array}
\usepackage{booktabs}
\usepackage{multirow}
\usepackage{graphicx}      %

\usepackage{amssymb}
\usepackage{pifont}
\usepackage{stackengine}

\usepackage{subcaption}
\usepackage{relsize} %
\usepackage{amsmath} %
\usepackage{float} %
\usepackage[accsupp]{axessibility}  %

\definecolor{iccvblue}{rgb}{0.21,0.49,0.74}
\usepackage[pagebackref,breaklinks,colorlinks,allcolors=iccvblue]{hyperref}
\usepackage{adjustbox}
\usepackage{graphicx}

\def\paperID{16} %
\def\confName{ICCV}
\def\confYear{2025}

\title{Express4D: Expressive, Friendly, and Extensible\\4D Facial Motion Generation Benchmark}

\author{Yaron Aloni\textsuperscript{1}, Rotem Shalev-Arkushin\textsuperscript{1}, Yonatan Shafir\textsuperscript{1}, Guy Tevet\textsuperscript{1}\\Ohad Fried\textsuperscript{2}, Amit Haim Bermano\textsuperscript{1}\\
\textsuperscript{1}Tel Aviv University \ \  \textsuperscript{2}Reichman University\\
{\tt\small yaronaloni@mail.tau.ac.il}
}

\begin{document}

\maketitle

\begin{abstract}
Dynamic facial expression generation from natural language is a crucial task in Computer Graphics, with applications in Animation, Virtual Avatars, and Human-Computer Interaction. However, current generative models suffer from datasets that are either speech-driven or limited to coarse emotion labels, lacking the nuanced, expressive descriptions needed for fine-grained control, and were captured using elaborate and expensive equipment. 
We hence present a new dataset of facial motion sequences featuring nuanced performances and semantic annotation. The data is easily collected using commodity equipment and LLM-generated natural language instructions, in the popular ARKit blendshape format. This provides riggable motion, rich with expressive performances and labels. We accordingly train two baseline models, and evaluate their performance for future benchmarking. Using our Express4D dataset, the trained models can learn meaningful text-to-expression motion generation and capture the many-to-many mapping of the two modalities. 
The dataset, code, and video examples are available on our webpage:\\
{\url{https://jaron1990.github.io/Express4D/}}
\end{abstract}
    
\section{Introduction}
\label{sec:intro}

\begin{figure}[t!]
\centering
\begin{overpic}[width=\columnwidth]{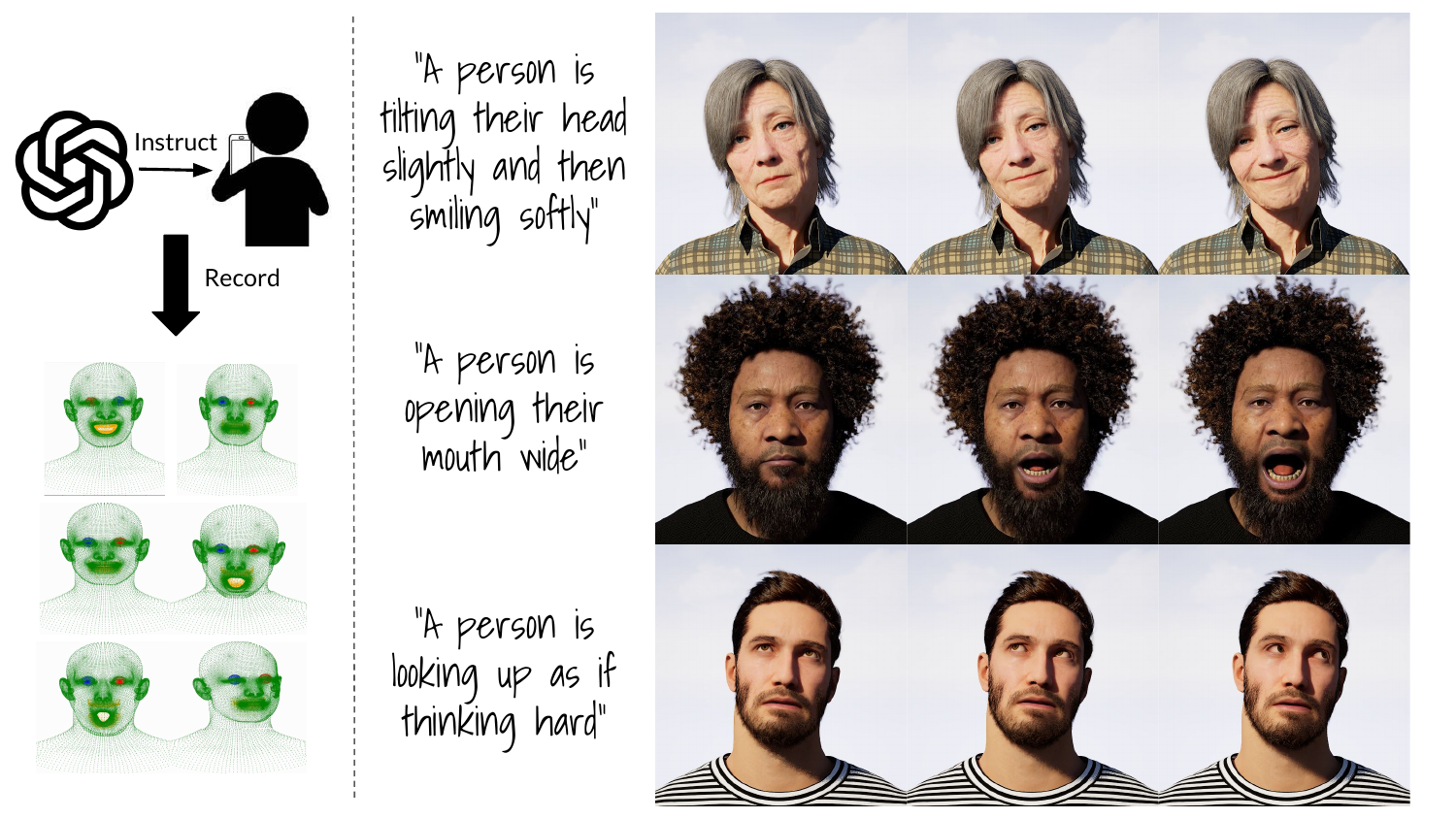}
\end{overpic}
\caption{
(Right) We construct our dataset, Express4D, by prompting a Large Language Model to generate a wide range of text instructions for facial motions. These instructions are then enacted by real users and recorded, resulting in an expressive, dynamic facial expression dataset. The dataset integrates seamlessly with animation tools, enabling rich and flexible character animation.
(Left) Three examples from Express4D are presented on different face ARKit rigs from the MetaHuman project by Epic Games \cite{metahumanMetaHumanHighFidelity}.
} 
\label{fig:teaser}
\end{figure}

Facial animation, and specifically dynamic facial expression generation, is one of the most sought-after disciplines of computer graphics since its emergence, with broad applications in entertainment, gaming, virtual reality, video conferencing, and human-computer interaction ~\cite{thambiraja2023imitator,kimmel2023let,zhang2025towards}. This elusive task demands subtle and nuanced precision, while perpetually carrying the risk of falling into the uncanny valley \cite{mori1970uncanny}.

In spite of its prevalence, automatically generating the subtle nuances and emotions has received limited attention in recent literature, where the main focus was put on text-to-speech applications \cite{VOCA2019, wang2021audio2head, Cheng_2018_CVPR}.
This is surprising since text-driven generation demonstrates impressive capabilities across various other domains, from high-quality images~\cite{rombach2021highresolution, gal2022textual, ho2020denoising} and videos~\cite{song2023consistency, blattmann2023videoldm, ma2025latte} to human body motion and dancing~\cite{tevet2023human, zhang2023generating, raab2024monkey}. Like many other fields employing machine learning, this stems largely from the scarcity of data. %

Existing facial motion datasets suffer from several critical limitations that hinder rich and semantic text-driven generation. Most datasets are either dedicated to speech~\cite{wu2023mmface4d,VOCA2019}, or continue the traditional animation approach of considering only a limited subset of categorical actions (e.g. happiness, anger, and disgust)~\cite{COMA:ECCV18,TMPEH:CVPR:2023}. While these subsets are commonly considered to span the range of human expressions, they do not capture subtle nuances. They hence might serve a human animator well through convenient decomposition, but they fail to represent 
richness of the facial
distribution and to provide corresponding semantics (see \cref{fig:tsne_and_count}~b). For example, `Happiness' could be expressed in many ways, and without semantic description, it is challenging to control the fine-grained details for generation of smiling softly while looking at a loved one, or winking and laughing.
Moreover, some facial motions express a combination of emotions, e.g. ``A person is transitioning from sadness to laughter as if remembering something funny'', or do not fit into a specific category, e.g. ``A person is looking up as if thinking hard''.
Hence, a dataset with a variety of facial motions derived from diverse detailed descriptions is needed.
A recent work in this direction~\cite{wu2024mmhead} estimates motion and expressions from monocular videos; however, this approach relies on automatic reconstructions, which again tends to fail in extracting exactly the same nuances, leading to critical information loss. 

To address these limitations, we introduce \emph{Express4D}, a 
facial motion dataset and baseline benchmark, comprising 1205 diverse sequences performed by 18 participants, featuring LLM-based natural language instruction labels which are performed and recorded; accompanied by two baseline models for benchmark comparisons. 
Our data is captured using off-the-shelf mobile apps~\cite{livelinkface}, leveraging commodity mobile camera and depth sensors in the popular ARKit~\cite{applearkit} blendshape format. This collection method provides precise 3D facial motion, but requires no additional equipment or complex lab settings. The used representation is compatible with common animation pipelines, facilitating the integration of generated animations in downstream tasks, in contrast to the common mesh-based datasets captured in elaborate lab settings. Together, this creates a comprehensive resource for text-to-motion research while maintaining compatibility with industry-standard animation pipelines. 
Due to the flexibility of our data representation, we can easily adapt the generated facial expressions to diverse characters (see \cref{fig:same_pose_diff_people}).
Our simple instruction generation pipeline and minimal equipment requirements further allow the easy extension of our dataset using personal devices.

We train two baseline text-to-facial motion generation models over our dataset 
that demonstrate the generation abilities of models trained on this data, and serve as a baseline for future reference.
Additionally, we train a facial motion feature extractor for automatic evaluations of facial motion generation, based on the common architecture used for evaluating human motion generation \cite{guo2022generating}.
We publicly release our code and dataset for further usage and exploration.

\section{Related Work}

\subsection{Face Representations}
The 3D Morphable Model (3DMM) \cite{blanz1999morphable}, was a seminal work in modeling 3D faces using PCA, modeling both shape and texture as linear combinations of principal components derived from 3D face scans. 
This foundational work enabled parametric control over facial identity and expression and laid the groundwork for many subsequent models \cite{egger20203d}.
Building on this idea, FLAME~\cite{FLAME:SiggraphAsia2017} introduced a parametric 3D face model that also employs PCA to represent facial shape and expressions using components learned from 3D scans of different identities and expressive performances. These components are also known as blendshapes.
Another prominent representation that emerged around the same time is Apple's ARKit blendshape model~\cite{applearkit}, which simplifies the approach by leveraging a fixed set of manually designed expression coefficients. ARKit enables efficient and robust real-time facial animation, making it especially suitable for deployment in consumer devices and virtual avatar systems.

\subsection{Face Motion Datasets}
Table \ref{tab:mm_dataset_comparison} summarizes prominent facial expression datasets and compares them.
While many datasets aim to capture the nature of human facial expressions, few include fine-grained annotations of the performed actions. Among the compared datasets, only MMHead~\cite{wu2024mmhead} and our \emph{Express4D} provide free-text descriptions. However, MMHead estimates these captions automatically based on recorded videos, whereas in Express4D, the expressions were performed in response to free-text prompts and captured with a depth camera. This makes Express4D the only dataset where expressions are both described in free language and intentionally enacted to match those descriptions.
Although ARKit is highly adopted by the industry, current facial motion datasets typically use either the FLAME facial representation~\cite{TMPEH:CVPR:2023,VOCA2019,wu2024mmhead,COMA:ECCV18,cosker2011facs} or the 3DMM mesh representation~\cite{Cheng_2018_CVPR,wu2023mmface4d}.
The recent DREAM-talk~\cite{zhang2023dream} estimates ARKit blendshapes from videos, while we collect our data using a depth camera.
Moreover, most datasets are either speech-driven~\cite{wu2023mmface4d,VOCA2019} or only contain categorical expression labels~\cite{COMA:ECCV18,TMPEH:CVPR:2023,zhang2023dream,cosker2011facs}.
In contrast, our dataset contains dynamic facial expressions driven by diverse free-form text.

\subsection{Human Motion Generation}
Human motion generation is the task of automatically synthesizing realistic human motions. Deep learning-based models~\cite{tevet2023human,zhang2023generating,guo2024momask,yi2023generating} typically learn from a dataset of motions represented by joint trajectories or body meshes (e.g. SMPL~\cite{SMPL:2015}).

Facial motion generation involves synthesis of facial movement over time, typically represented as 3D landmarks, blendshape coefficients, or mesh deformations. 
Two common approaches are speech-driven and expression- or emotion-driven generation. Speech-based methods~\cite{fried2019text,yao2021talkinghead,VOCA2019,xing2023codetalker,fan2022faceformer} aim to generate lip movements and facial gestures synchronized with spoken audio, often using audio-to-motion models. Expression-based methods focus on generating facial dynamics that convey specific emotions or mimic natural expressions, either from speech~\cite{peng2023emotalk}, semantic labels~\cite{zhang2023dream,TMPEH:CVPR:2023}, or other modalities~\cite{wu2024mmhead}. 
Recent advances leverage deep generative models and neural rendering to produce high-fidelity and temporally coherent results. 

\section{Express4D}
\begin{figure}[t!]
  \centering
    \includegraphics[width=\columnwidth]{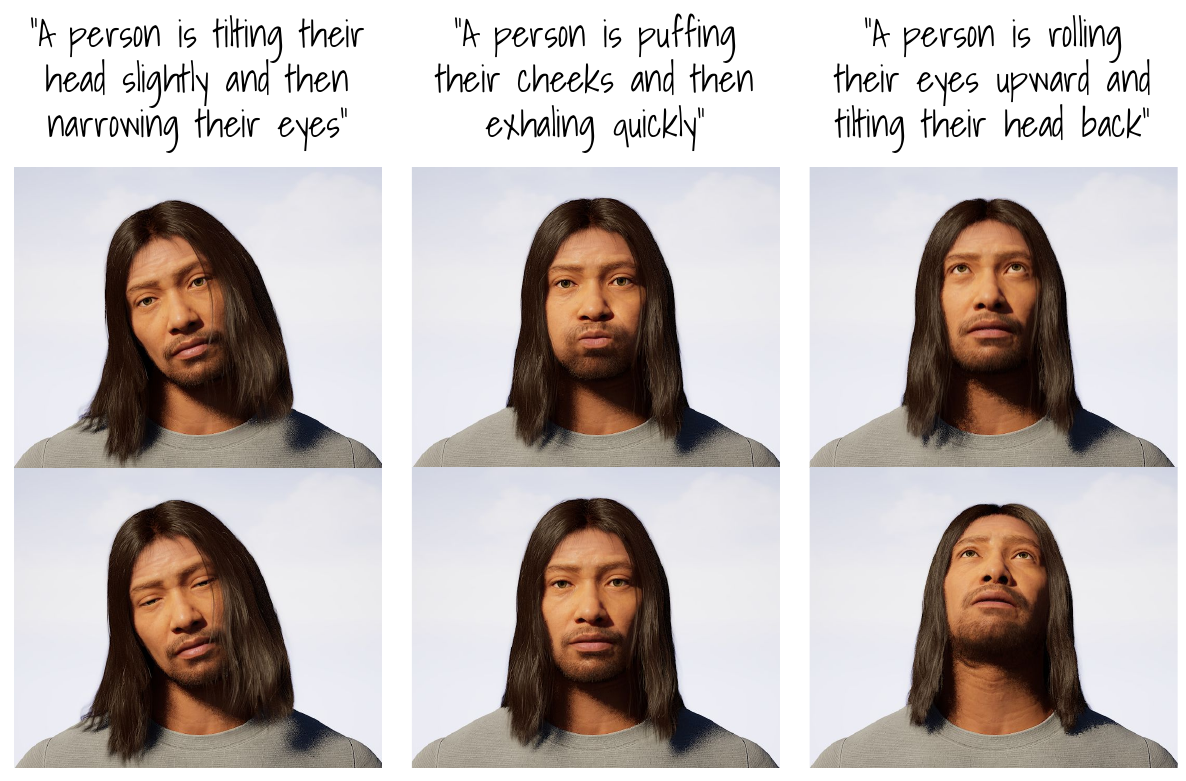}
   \caption{Our dataset contains diverse facial motions. Some motions can be easily expressed via text but are not easily categorized.}
\label{fig:diff_pose_same_person}
\end{figure}

\begin{figure}[t!]
  \centering
    \includegraphics[width=\columnwidth]{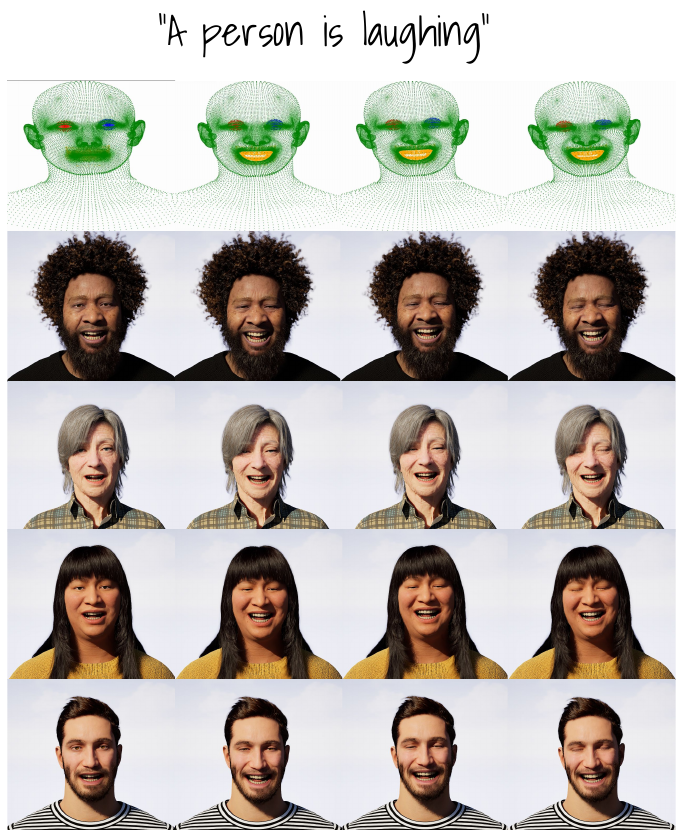}
   \caption{The blendshapes representation of our data enables easy transfer of facial motions between different characters.
The top row presents the vanilla ARKit blendshape representation, followed by the same motion enacted by different ARKit rigs from the MetaHuman project by Epic Games \cite{metahumanMetaHumanHighFidelity}.}
\label{fig:same_pose_diff_people}
\end{figure}

We constructed the \emph{Express4D} dataset to enable automatic facial motion generation from natural language. It consists of diverse facial motions (see \cref{fig:diff_pose_same_person}) that are riggable and can be easily transferred to different characters (see \cref{fig:same_pose_diff_people}).
Each data point consists of a short text prompt describing a facial expression or reaction, paired with a dynamic sequence of 3D facial motion.

\subsection{Dataset Collection}
\label{sec:collection}
\begin{figure*}[htp]
\centering
\begin{overpic}[width=1\linewidth]{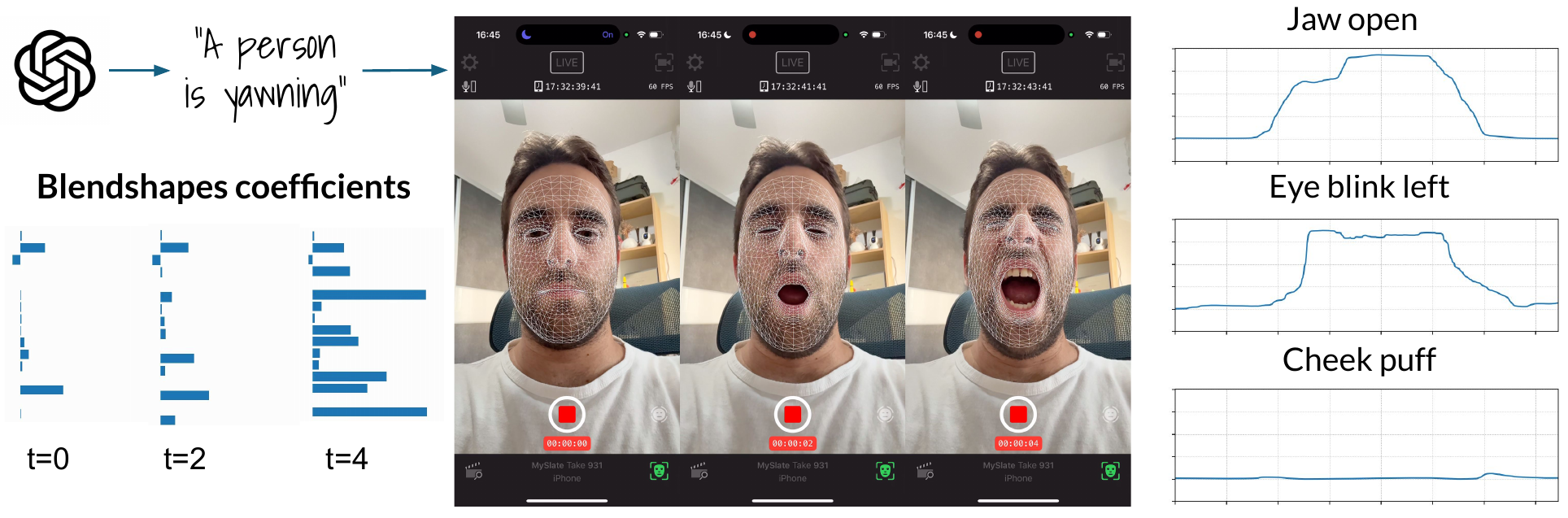}
\end{overpic}
\caption{
Data collection overview. We start by querying GPT to output text prompts of faciel expressions. Then, a person captures themselves performing the required prompt using the Live Link Face app on an iPhone. The app uses the depth camera of the iPhone to estimate the ARKit blendshapes coefficients at each frame. Each blendshape coefficient varies throughout the video in compatible to the performed motion. For example, when a person is yawning, both the jaw opens and the eyes close gradually throughout the video, and the cheeks are not puffed, so the coefficients of these blendshapes will be low or high accordingly.
}
\label{fig:overview}
\end{figure*}

As illustrated in \cref{fig:overview}, we start by collecting textual descriptions of dynamic facial expressions using an LLM, and continue with performing and capturing the required facial motions. In the following sections, we elaborate on each of these parts.

\subsubsection{Text Generation}
To generate diverse and expressive prompts, which are short descriptions of facial expressions in natural language, covering a wide emotional and behavioral range (e.g., ``A person is raising one eyebrow and smirking", ``A person is giggling", ``A person is puffing their cheeks and blinking rapidly", etc.),
we employed in-context learning (ICL) ~\cite{brown2020language}
by querying ChatGPT-4o~\cite{achiam2023gpt} with the following prompt, which includes illustrative examples:

\begin{table*}[htp]
\centering
\setlength{\tabcolsep}{3pt}
\renewcommand{\arraystretch}{1.2}

\resizebox{\textwidth}{!}{
\begin{tabular}{l|ccc|ccc|cc|c}
\hline
\multirow{2}{*}{\textbf{Dataset}} & 
\multicolumn{3}{c|}{\textbf{Acquisition}} & 
\multicolumn{3}{c|}{\textbf{Annotation}} & 
\multicolumn{2}{c|}{\textbf{Scale}} & 
\multirow{2}{*}{\textbf{Representation}} \\

& Env. & Tech. & Real/Gen. & Class & Transc. & Text & Subj. & Dur. (min) &\\
\hline
\hline

BU-4DFE~\citep{6553788}          & Lab & 3 cameras & R & \checkmark & -- & -- & 101 & 40 & Mesh \\

BP4D+~\citep{zhang2016multimodal} & Lab & 4 cameras & R & \checkmark & -- & -- & 140 & 777 & Mesh\\ 

4DFAB~\citep{Cheng_2018_CVPR}    & Lab & 6 cameras & R & \checkmark & \checkmark & -- & 180 & 509 & Mesh \\

VOCASET~\citep{VOCA2019}         & Lab & 6 cameras & R & -- & \checkmark & -- & 12 & 29  & Mesh (FLAME)\\ 
COMA~\citep{COMA:ECCV18}          & Lab & 12 cameras & R & \checkmark & -- & -- & 12  & 5.7 &  Mesh (FLAME) \\
FaMoS~\citep{TMPEH:CVPR:2023}    & Lab & 24 cameras & R & \checkmark & -- & -- & 95 & 166 & Mesh (FLAME)\\ 
MMFace4D~\citep{wu2023mmface4d}  & Lab & 3 cameras & R & \checkmark & -- & -- & 431  & 2166  & Mesh (3DMM) \\
Florence4D~\citep{https://doi.org/10.48550/arxiv.2210.16807} & Mix & \makecell{3dMD scanner\\ unspecified setup} & Both  & \checkmark & -- & -- & 95 & 5075 & Mesh (FLAME) \\ 
D3D-FACS~\citep{cosker2011facs}  & Lab & 8 cameras & R & \checkmark & -- & -- &  10 & 65 & Mesh \\ 
\hline
MMHead~\citep{wu2024mmhead}   & --  & Mono. Rec. & R & \vix & \vix & \vix & N/A & 2940 & 56 Blendshapes (FLAME)\\ 
\textbf{Express4D (Ours)} & \textbf{Commodity} & \textbf{1 Cell phone} & \textbf{R} & -- & -- & \textbf{\checkmark} & \textbf{18} & \textbf{89.6} & \textbf{61 Blendshapes (ARKit)}\\

\hline
\end{tabular}
}

\caption{Comparison of multimodal facial datasets. \checkmark indicates availability, \vix indicates data annotated by model, and -- means not available.}
\label{tab:mm_dataset_comparison}
\end{table*}

\begin{figure}[htpb]
\centering

\subcaptionbox{}[\columnwidth]{%
\includegraphics[width=\linewidth]{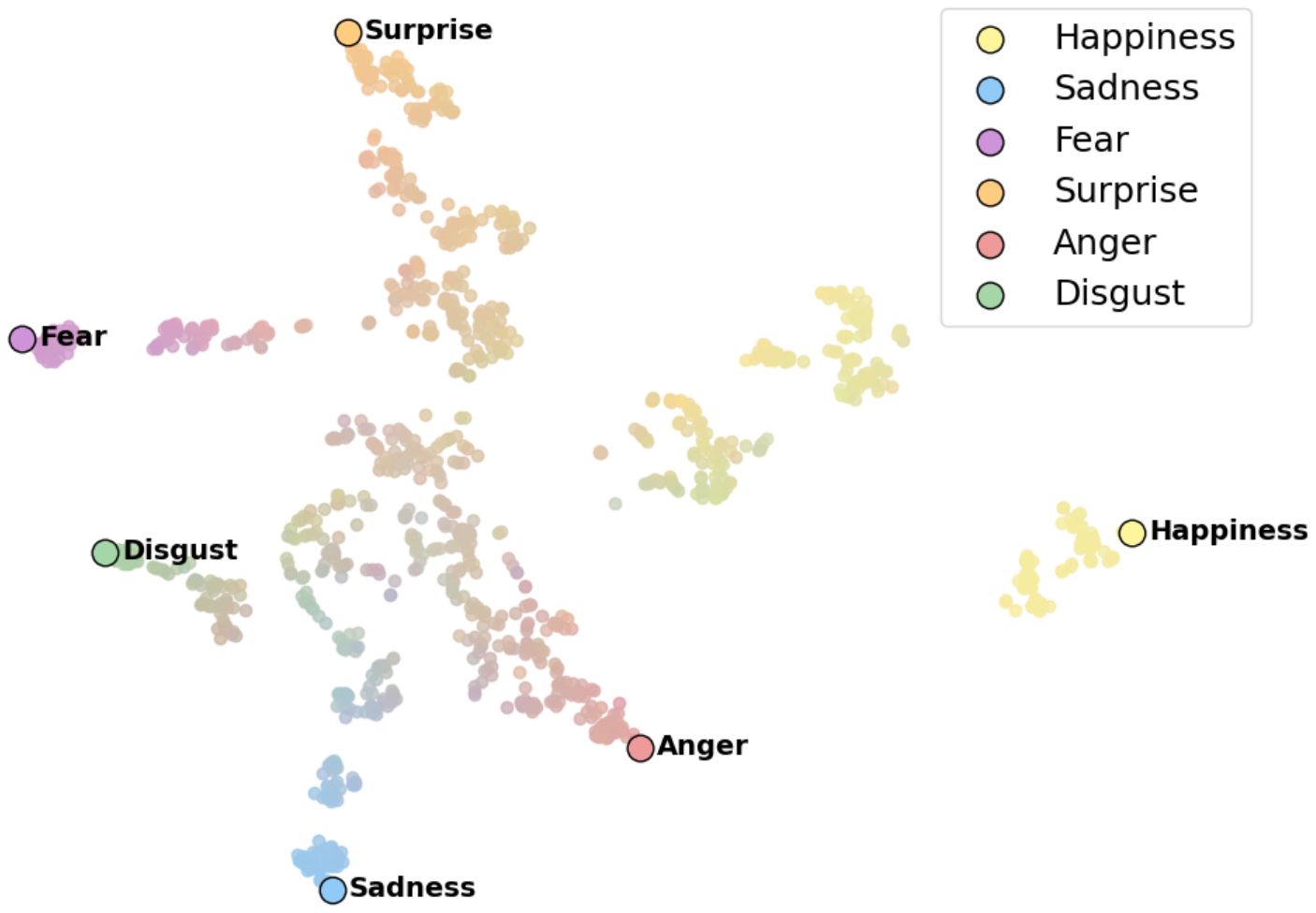}}

\hfill
\subcaptionbox{}[\columnwidth]{%
\includegraphics[width=\columnwidth]{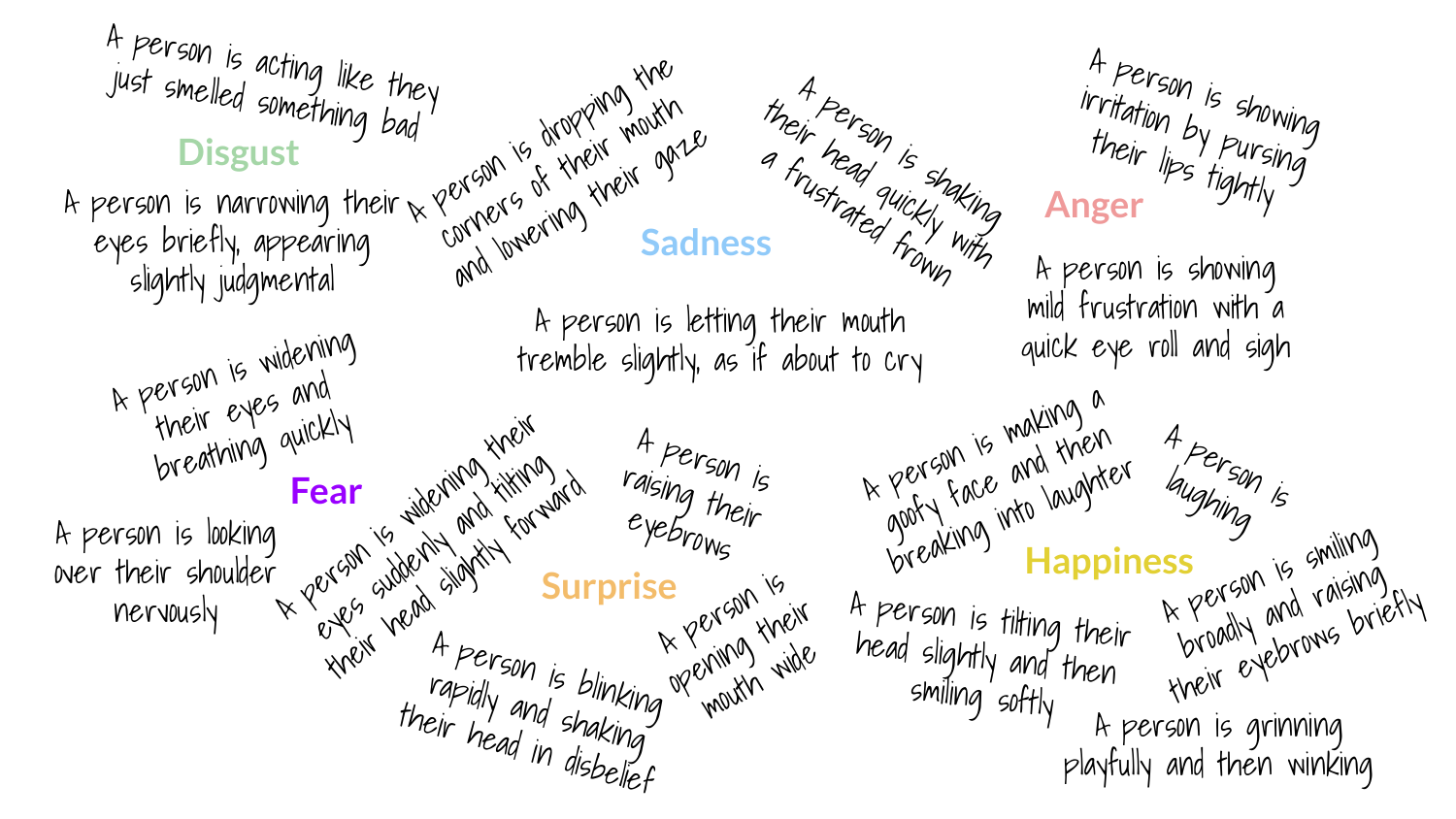}}

\hfill
\subcaptionbox{}[\columnwidth]{%
\includegraphics[width=\columnwidth]{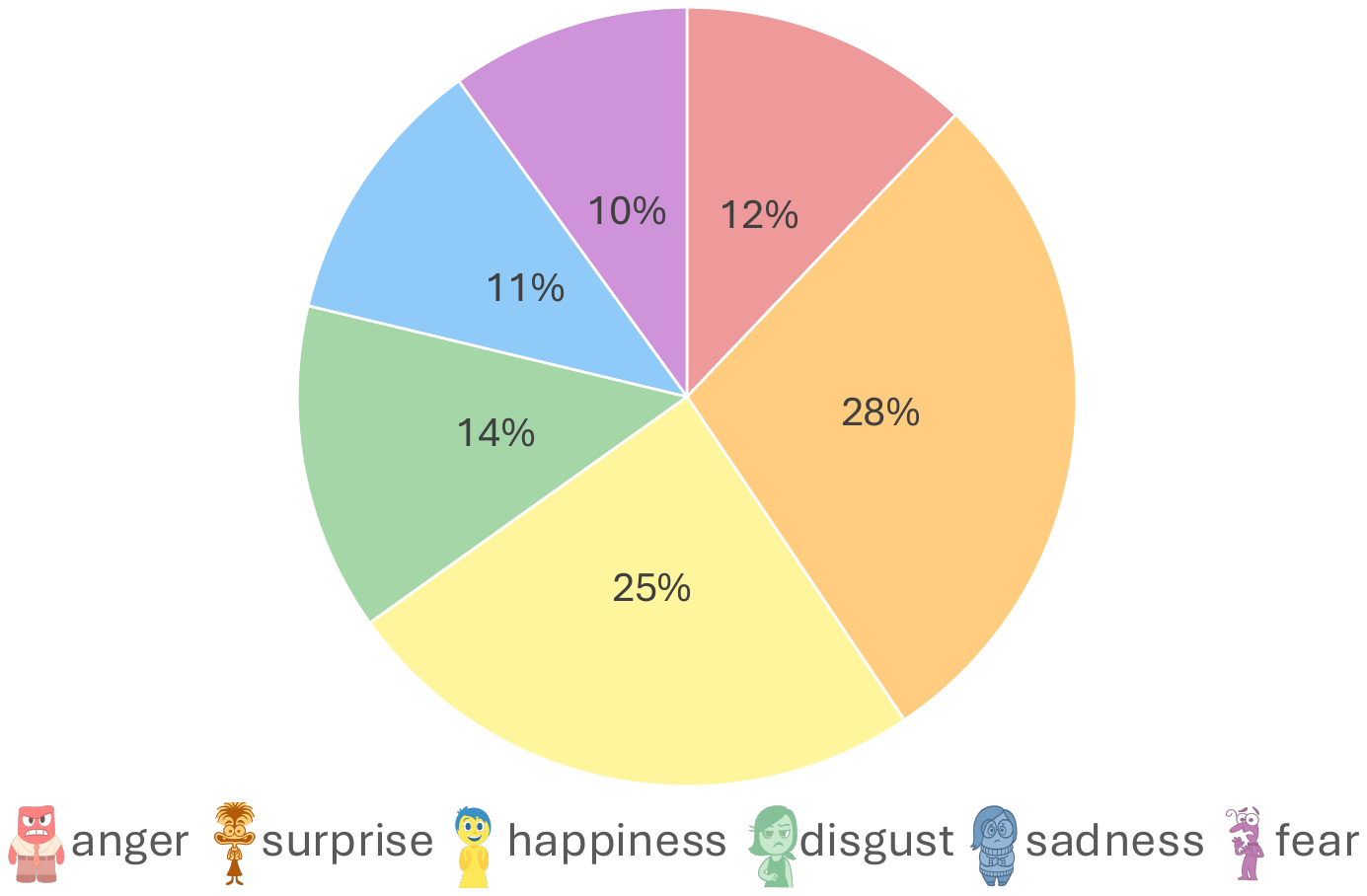}}

\caption{Dataset analysis: (a) t-SNE projection of the texts in our dataset, split into their matching FACS~\cite{ekman1978facial} categories by GPT. As some texts belong to multiple categories, the colors represent the blend of categories to which each text belongs. 
(b) Semantic grouping example -- each FACS category is diverse and contains many different expressions, and vice versa -- sentences might fit in numerous different categories, e.g. ``A person is widening their eyes suddenly and tilting their head slightly forward" could mean both surprise and fear. 
(c) Emotion label proportions. The proportions of texts in \emph{Express4D}, based on the highest scored FACS category from the text categorization.}
\label{fig:tsne_and_count}
\end{figure}

\textit{``I need a list of 100 face expressions prompts for a person to perform in front of a camera. e.g: ``a person smiling to the camera while moving his head left to right". The actors are not professional so the prompts should be not so hard to perform. Also - this prompts will be used, coupled with the video - in order to train a text to expression generative model.''} 

The dataset was built incrementally in batches of 100 samples, allowing manual inspection of the resulting sequences and iterative refinement of the prompt generation process. During iterative refinement, the prompts were progressively modified to include underrepresented expressions and more varied emotional content.
For example, after observing that initial prompts often produced simple or isolated expressions, we adapted the generation prompt to explicitly request combinations of expressions \textit{(e.g., “A person looks right and then smiles” or “A person smiles with teeth and then raises their eyebrows”)}, thereby encouraging more dynamic and varied facial motions in subsequent batches.
Using an LLM instead of manually crafting textual prompts for actors helped increase diversity across prompts. In our iterative process with chatGPT we explicitly asked it to avoid repetitions and encouraged diversity, which lead to rich and diverse expression instructions. \\
Prompting chatGPT for new instructions is fast and easy, which makes the overall data collection process more scalable and efficient.

\subsubsection{Motion Collection}
\label{subsec:motion_collection}

The motions were captured using the Apple ARKit framework~\cite{applearkit}, which 
provides real-time facial tracking through an iPhone’s TrueDepth camera system.
To enable scalable and accessible data collection, we used the Live Link Face app~\cite{livelinkface} by Epic Games on an iPhone 16 Pro, which leverages its front-facing TrueDepth sensor to extract a total of 61 blendshape coefficients per frame. These include 52 facial expression coefficients, 3 head rotations, and 6 eye rotations. The data was recorded in real time at 60Hz, providing dynamic expression sequences suitable for motion learning and generation.
The app estimates semantically meaningful blendshapes, as well as head and eye rotations, and offers several practical advantages. 
First, it enables easy and low-cost data acquisition using widely available consumer devices. This makes the dataset collection process highly scalable and allows other researchers or creators to extend the dataset independently using their own devices (see \cref{sec:conclusion}). 
Second, it provides a standardized set of interpretable blendshape coefficients with semantically meaningful names \textit{(e.g., jawOpen, browDownLeft)}, making the data more accessible for editing, analysis, and retargeting across animation pipelines.

Participants performed each prompt under consistent lighting across sessions. The subject was instructed to remain in a frontal position without translating their head in space, but was allowed to rotate it and use full facial expressivity.
Each session began with calibration using the Live Link Face app to align the face mesh and stabilize the mapping of blendshape coefficients. This step helped mitigate variability due to lighting conditions, facial posture, or transient changes such as eye openness that might occur across different recording days.
In each recording session, the built-in functionality of the Live Link Face app was used to record ARKit blendshape data locally on the device. 

\subsubsection{Data Collection Website}
\label{subsec:ui}
\begin{figure}[t!]
  \centering
    \includegraphics[width=\columnwidth]{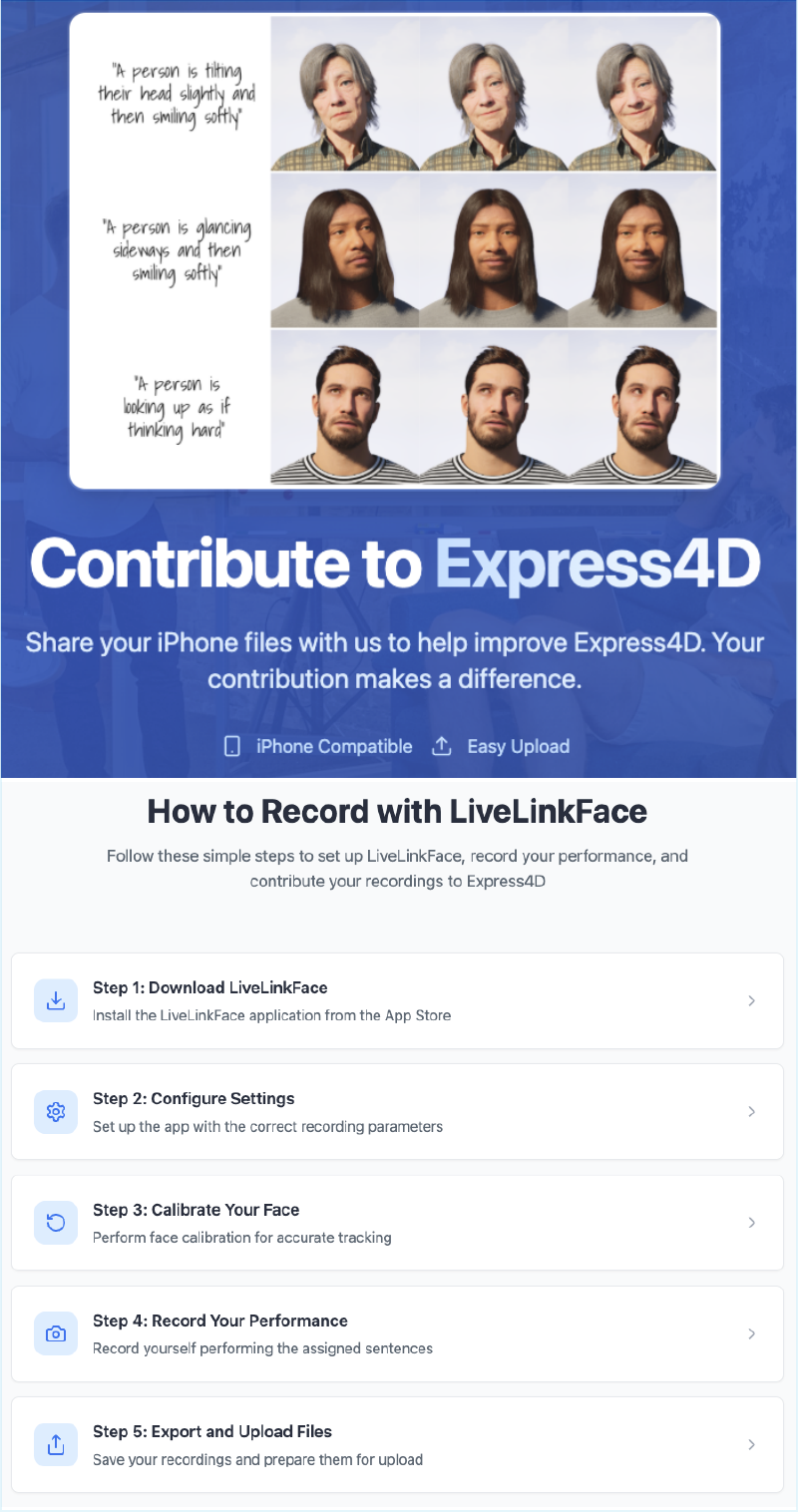}
   \caption{We provide a WebUI for crowd-sourced data collection via personal iPhone cameras, which allows easy extensibility of Express4D by the community.}
\label{fig:website}
\end{figure}

We release a user interface (UI) that enables crowd-sourced data collection via personal devices, allowing the community to contribute and expand the Express4D dataset. The UI provides clear instructions for self-recording, as illustrated in \cref{fig:website}. 
Each contributor will be assigned batches of 10 sentences to perform, along with guidance for calibration, recording, and uploading. All submitted data will be reviewed to filter out low-quality or disruptive contributions, and validated sequences will be published on a regular basis. Link to the UI can be found on our \href{https://jaron1990.github.io/Express4D/}{project page}.

\subsection{Data Format and Features}

The \emph{Express4D} dataset comprises 1205 sequences.
Each data sample consists of an English free-form text prompt, paired with a dynamic sequence of facial motion captured using ARKit.
Every motion frame is represented as a 61-dimensional vector, as described in \cref{subsec:motion_collection}. Head and eye rotations are in Euler angles (roll, pitch, yaw).
Blendshape names conform to the ARKit specification, enabling fine-grained control and animation. Sequences are recorded at 60FPS and range from 1.75 to 10 seconds in duration, depending on the prompt complexity and performance.
The data is stored in a lightweight and training-friendly format (CSV), facilitating easy loading and preprocessing. All coefficients remain in ARKit scale without normalization, preserving their native distribution in the dataset. 
Each CSV file begins with a timecode column (in seconds) followed by the 52 facial blendshape coefficients and the 9 rotation values. The blendshape names follow the Apple ARKit specification (e.g., \texttt{jawOpen}, \texttt{mouthSmileLeft}, \texttt{eyeBlinkRight}), ensuring compatibility with standard AR/VR and game development tools.

The streamlined format of \emph{Express4D} supports low-overhead integration into training pipelines and enables flexibility in exploring both direct blendshape regression and intermediate representation approaches.

\subsection{Dataset Analysis}

Our \emph{Express4D} dataset contains 1205 facial motion sequences, performed by 18 participants. It is designed to span the space of natural expressions through diverse textual descriptions. 
The widely-adopted Facial Action Coding System (FACS) framework~\cite{ekman1978facial}, which serves as the gold standard for facial expression analysis, divides facial expressions into six basic emotional categories (happiness, sadness, anger, fear, surprise, and disgust). Following FACS, we analyze the distribution of our collected expressions across these categories. 
To analyze expression distribution across FACS categories, we queried ChatGPT-4o to return a score vector summing to 1, which represents how much each sentence in our dataset belongs to each of the categories of FACS. 
e.g -- for the prompt \textit{"A person is laughing naturally as if hearing a funny joke."} and the categories [\textit{happiness, sadness, fear, surprise, anger, disgust}] GPT answered with the vector {[\textit{\textbf{0.812}, 0.034, 0.021, 0.067, 0.019, 0.047}]} respectively, with \textit{happiness} clearly dominating. In contrast, the prompt \textit{"A person is pressing their lips together firmly and glancing sideways nervously"} yielded {[\textit{0.042, 0.067, \textbf{0.312, 0.184, 0.231, 0.164}}]}, where \textit{fear} received the highest score, but \textit{surprise, anger} and \textit{disgust} also showed substantial contributions, reflecting the prompt’s ambiguity and emotional complexity.
We confirmed the analysis using other LLMs (Claude-3~\cite{claude2024}, Gemini-2.5~\cite{google2025gemini}) with consistent results.

\cref{fig:tsne_and_count} (a) shows the distribution of prompts across FACS-based emotional groupings, demonstrating coverage across the expression spectrum despite not being constrained by these categories during collection.
Prompts are clustered by general emotional category while maintaining internal variation that reflects the subtle differences expressible only through natural language. Their colors represent the blend of categories to which each sentence belongs. 
As illustrated, categorical labels alone fail to capture the rich nuances present in our dataset. Unlike traditional datasets that rely on discrete emotional categories, our natural language prompts offer fine-grained distinctions within emotion classes. For example, within the "happiness" category, our prompts distinguish between simple expressions like "A person is laughing" and more complex, dynamic sequences such as "A person is grinning playfully and then winking." This granularity is illustrated in our \mbox{t-SNE} visualization (\cref{fig:tsne_and_count} (a)) and exemplified in \cref{fig:tsne_and_count} (b).
\cref{fig:tsne_and_count} (c) presents sentence distribution by the dominant FACS category per prompt.

\section{Experiments}
\begin{table}[t!]
\centering
\resizebox{\columnwidth}{!}{
\begin{tabular}{lcccccc}
\toprule
\multirow{2}{*}{\textbf{Model}}&
\multirow{2}{*}{\textbf{FID $\downarrow$}} & \multicolumn{3}{c}{\textbf{R-Precision $\uparrow$}} & \multirow{2}{*}{\textbf{Diversity $\rightarrow$}}& \multirow{2}{*}{\textbf{\makecell{Multimodal\\Distance} $\downarrow$}}  \\
\cline{3-5}
 & & top-1 & top-2 & top-3 & & \\
\midrule
GT          & 0.157 & 0.335 & 0.488 & 0.585 & 7.810 & 4.248 \\
\hline
MDM        & 1.705 & 0.251 & 0.388 & 0.480 & 8.370 & 4.917 \\
T2M-GPT     & 1.897  & 0.293  & 0.428  & 0.528  & 7.763  & 6.809  \\
\bottomrule
\end{tabular}
}
\caption{Quantitative comparison of text-to-facial-motion generation models on the Express4D test set. 
$\rightarrow$ indicates that better is closer to ground-truth performance.
}
\label{tab:comparison}
\end{table}

The Express4D dataset, combined with adaptations of the popular HumanML3D~\cite{guo2022generating} metrics for body motion generation, comprises a benchmark for text-to-dynamic facial expressionemotion generation. 
In thisis section, we start by describing ourthe evaluation metrics (\ref{subsec:eval_model}), and then provide two baseline models, trained over our data, based on the state-of-the-art models for human body motion generation: diffusion models and VQ-based models (\ref{sec:MDM}).

\subsection{Evaluation}
\label{subsec:eval_model}
To evaluate the baseline models trained over our dataset, we adopt the common evaluation metrics used by text-to-motion works~\cite{tevet2023human,zhang2023generating,guo2022generating};
Fréchet Inception Distance (FID), R-precision, diversity, and multimodal distance.
FID evaluates the realism of the generated motion, by comparing the statistics of the generated motions with those of the real data.
R-precision and multimodal distance evalute the alignment between the generated motion and the text prompt.
To calculate FID and R-precision over our data, we train a feature extractor using the architecture suggested in HumanML3D~\cite{guo2022generating}, which is the commonly used evaluation model for text-to-motion. This model learns to map similar text and motion sequences to feature vectors that are geometrically close to each other.

\subsubsection{Fréchet Inception Distance (FID)}
FID measures the quality of generated motions by comparing the distribution of generated sequences to real sequences in the learned feature space. Specifically, we compute the Fréchet distance between two multivariate Gaussian distributions fitted to the real and generated motion features:

\begin{figure}[t!]
  \centering
    \includegraphics[width=\columnwidth]{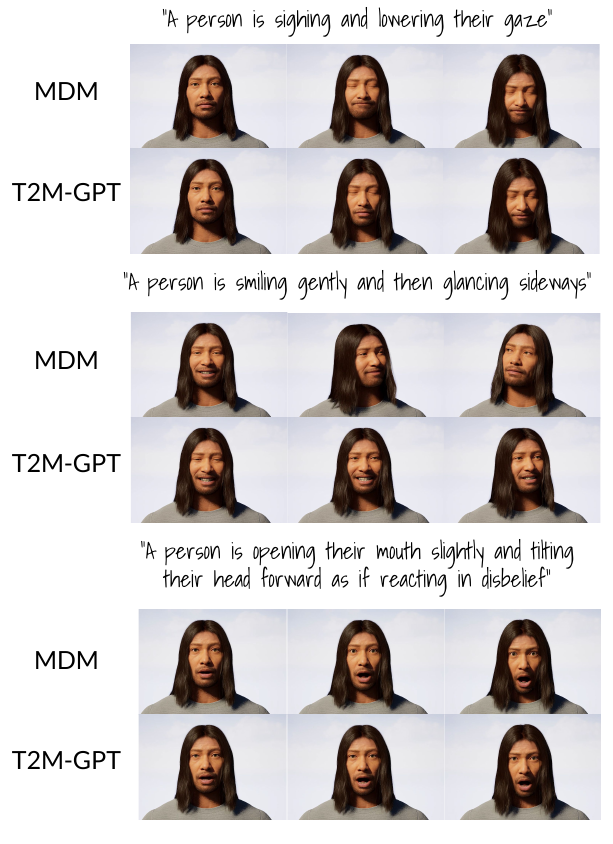}
   \caption{
   Generation results of MDM and T2M-GPT trained over our dataset. Both models generate realistic and plausible results that align with the required text prompts.  
   }
\label{fig:gen_res}
\end{figure}

$FID = ||\mu_r-\mu_g||_2^2 + Tr(\Sigma_r-\Sigma_g-2(\Sigma_r\Sigma_g)^{\frac{1}{2}})$\\
where $ \mu_r, \Sigma_r, \mu_g, \Sigma_g$ are the mean and covariance of real and generated motion features, respectively. 
Lower FID scores indicate that generated motions are more similar to real motions in terms of feature distribution.

\subsubsection{R-Precision}
R-precision evaluates the semantic alignment between generated motions and their corresponding text descriptions. We compute this metric using batches of 32 text-motion pairs. For each generated motion in the batch, we calculate the cosine similarity between the motion's feature vector and all 32 text feature vectors in the batch. R-precision is the percentage of cases where the ground truth text description ranks in the top R positions based on similarity scores. We report R-precision for $R \in \{1, 2, 3\}$, where higher values indicate better text-motion correspondence.

\subsubsection{Diversity}
Diversity quantifies the variability in generated facial motions across different expression types. To compute this metric, we randomly sample two subsets of equal size (Sd = 48) from all generated motions and extract their corresponding feature representations $\{v_1, ..., v_{Sd}\}$ and $\{v'_1, ..., v'_{Sd}\}$. The diversity score is calculated as the average distance between corresponding features from the two subsets:

$Diversity = \dfrac{1}{S_d} \displaystyle\sum_{i=1} ^{S_d} ||v_i-v'_i||_2$ \\
Higher diversity scores indicate greater variation in the generated motions, demonstrating that the model avoids mode collapse and produces a rich variety of facial expressions.

\subsubsection{Multimodal distance}
Multimodal distance measures the semantic alignment between text and motion modalities by computing the average distance between text and motion features in the learned embedding space:

$ MModDist = \dfrac{1}{N} \displaystyle\sum_{i} ||f_{text}(t_i) - f_{motion}(m_i)||_2$ \\
where $f_{text}$ and $f_{motion}$ are text and motion encoders respectively, $t_i$ is the text description, and $m_i$ is the corresponding motion sequence. Lower multimodal distance indicates better alignment between text descriptions and generated motions.

\subsection{Text-to-Expression Baseline Models}
\label{sec:MDM}

To evaluate the generation ability of facial motions based on our data, we train two models over 704 facial sequences from our dataset, based on the current most common approaches for text-to-human motion generation; a diffusion transformer (e.g. MDM~\cite{tevet2023human}), and a VQ-VAE (e.g. T2M-GPT~\cite{zhang2023generating}). The training code, along with the models checkpoints, is publicly available on our \href{https://jaron1990.github.io/Express4D/}{project page}.

Although we applied a straightforward adaptation in both cases, we observe that the two baselines are capable of modeling the intricate facial motion distribution as reflected by Express4D and generating high-fidelity motions accordingly.
Generation examples are presented in \cref{fig:gen_res} and in our \href{https://jaron1990.github.io/Express4D/}{project page}. Both models produce plausible results that align with the given text prompts.

We evaluate the ability of both models to generate new facial motions given a free-text quantitatively using our pre-trained evaluator and present the results in \cref{tab:comparison}.
When trained over our dataset, the MDM architecture yields slightly better FID and Multimodal Distance scores, whereas the T2M-GPT architecture yields slightly better R-Precision and diversity results. However, both architectures yield promising results with realistic dynamic facial expressions that align with the textual descriptions.

\subsection{Limitations}
\label{sec:lim}
\begin{figure}[t!]
  \centering
    \includegraphics[width=\columnwidth]{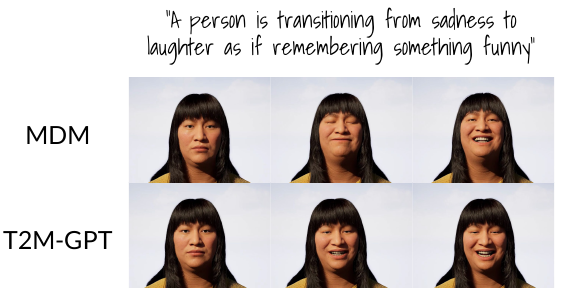}
   \caption{Generation limitations. Models struggle to generate complex text prompts that include multiple different facial expressions.}
\label{fig:lim}
\end{figure}

\begin{itemize}
    \item \textbf{Generation limitations:} Although both human motion generation architectures succeed in generating realistic facial expressions that align with the given text prompts, they have some failure cases.
Both models struggle with generating complex motions that involve multiple different expressions one after the other, e.g., "A person is transitioning from sadness to laughter as if remembering something funny". In that case, they often generate only one of the required expressions (see \cref{fig:lim}). This stems from limitations in the underlying architectures which often struggle to generate a sequence of motions.

\item \textbf{Data limitations:} The participants who performed the facial expressions in our dataset are not professional actors. As such, the expressions are more natural. However, some expressions that are hard to produce spontaneously (e.g., crying) might be missing from the data. 
Additionally, although the dataset includes motions from 18 different participants, it may not fully encompass the diversity of facial appearances across the broader population and from different ethnicities.
However, as our dataset can be easily extended via our webUI (see \cref{subsec:ui}), we invite future work to extend the dataset with additional examples from different people, including professional actors.
\end{itemize}

\section{Broader Impact}

Text-to-dynamic expression generation is an important task with a wide range of applications, including gaming, video conferencing, and human-computer interaction. These applications can benefit from more expressive and natural communication enabled by generated facial dynamics with fine-grained control.

However, we recognize the potential for misuse of this technology, particularly in the creation of deep-fake content, and we strongly oppose such use. To address this concern, we release our dataset under a license that explicitly prohibits malicious applications, including deep-fake generation. 
In addition, we are actively developing tools to identify and flag deep-fake content in order to prevent misuse of such generative technologies \cite{agarwal2020detecting,knafo2022fakeout,sinitsa2024deep}.

\section{Conclusion}
\label{sec:conclusion}

We introduced \emph{Express4D}, a dataset of dynamic facial expressions annotated with free-form text. By moving beyond predefined emotion categories, our approach supports more nuanced and controllable facial motion generation. The data was collected using personal mobile devices in an ARKit-compatible format, making it easy to extend and integrate into existing animation workflows.

Express4D is more than a dataset: it is also a benchmark. We adapted popular text-to-motion evaluation metrics to the facial domain and trained two baseline models based on state-of-the-art architectures. These baselines provide a starting point for future work, and we hope Express4D will serve as a common ground for developing and comparing facial motion generation models.
As a potential direction for future work, we suggest leveraging vision-language models (VLMs) to automatically validate user-contributed sequences by assessing their semantic alignment with the associated prompts.

While much of the current research focuses on speech-driven facial animation, expressive motion is equally important for realism and emotional fidelity. We show that natural language provides a more flexible and descriptive way to guide generation, capturing subtleties that discrete emotion labels cannot.

To support further progress, we release a web interface designed to expand Express4D. It uses an LLM to generate expressive prompts, which can be acted and recorded using personal devices. While its primary goal is to extend the dataset, we also see it as a valuable platform for collecting future multimodal data that combines speech and expression.
We hope Express4D will be a practical, extensible resource for the community and a solid foundation for future research in facial motion generation.

\newpage
\section*{Acknowledgments}
We thank Or Lichter for his useful suggestions and references.
This research was supported in part by Len Blavatnik and the Blavatnik Family Foundation, the Deutsch Foundation, a scholarship from the Center for AI and Data Science at Tel Aviv University (TAD) with the support of Dr. Monique Barel, and ISF grant numbers 1337/22 and 1574/21.

{
    \small
    \bibliographystyle{ieeenat_fullname}
    \bibliography{main}
}

\end{document}